\documentclass[aps,showpacs,showkeys]{revtex4}

\usepackage{amsmath}

\begin{document}


\newcommand\beq{\begin{equation}}
\newcommand\eeq{\end{equation}}
\newcommand\bea{\begin{eqnarray}}
\newcommand\eea{\end{eqnarray}}
\newcommand\ket[1]{|#1\rangle}
\newcommand\bra[1]{\langle#1|}
\newcommand\sd[2]{$\mathcal{#1}_#2$}

\newcommand\jo[3]{\textbf{#1}, #3 (#2)}


\title{\Large\textbf{Controlled Quantum Secret Sharing}}

\author{\bf Chi-Yee Cheung}

\email{cheung@phys.sinica.edu.tw}

\affiliation{Institute of Physics, Academia Sinica\\
             Taipei, Taiwan 11529, Republic of China}


\begin{abstract}

We present a new protocol in which a secret multiqubit
quantum state $\ket{\Psi}$ is shared by $n$ players and $m$
controllers, where $\ket{\Psi}$ is the encoding state of a
quantum secret sharing scheme. The players may be
considered as field agents responsible for carrying out a
task, using the secret information encrypted in
$\ket{\Psi}$, while the controllers are superiors who
decide if and when the task should be carried out and who
to do it. Our protocol only requires ancillary Bell states
and Bell-basis measurements.

\end{abstract}



\maketitle


Cryptography is the art and science of concealing a secret
message from unauthorized parties. At present the most
widely used cryptographic system is the RSA public-key
protocol invented by Rivest, Shamir, and Adleman in 1978
\cite{RSA78}. However the security of this protocol is not
proven, other than the fact the it is very hard to crack
with our present knowledge of mathematics and technology.

The idea of quantum cryptography was first proposed in the
1970's by Wiesner \cite{Wiesner83}. In recent years, the
field of quantum key distribution(QKD) has found many
fruitful applications of quantum information theory
\cite{Gisin-02}. Moreover, secure distributions of secret
cryptographic keys have been demonstrated in and outside
scientific laboratories
\cite{Bennett-92,Muller-96,Ribordy-01,Hughes-02}. The first
provably secure QKD protocol was constructed in 1984 by
Bennett and Brassard \cite{BB84} using polarized single
photons. Quantum entanglement assisted QKD was first
proposed by Ekert in 1991 \cite{Ekert91}. Since then, many
other QKD protocols have appeared in the literature.

In 1979 Blakely \cite{Blakely79} and Shamir \cite{Shamir79}
introduced the notion of secret sharing as a means of
safeguarding cryptographic keys. The idea is as follows.
Suppose Alice wants to send a secret message to a remote
location, and she has a choice of sending it to either
agent Bob or agent Charlie. In order to reduce the risk of
possible leakage and misuse of the message, it is often
safer for her to split the message into two shares and send
them to Bob and Charlie separately, such that either one
alone has absolutely no knowledge of the message. Bob and
Charlie can reconstruct the original secret message if and
only if they cooperate with each other. More generally, in
a so-called $(k,n)$-threshold scheme, the secret is divided
into $n$ shares, such that any $k$ of those shares can be
used to reconstruct the secret, while any set of less than
$k$ shares contains absolutely no information about the
secret at all.

Quantum secret sharing (QSS) refers to the implementation
of the secret sharing task outlined above using quantum
mechanical resources. Hillery \textit{et al.}
\cite{Hillery-99} and Karlsson \textit{et al.}
\cite{Karlsson-99} were the first to propose QSS protocols
using respectively three-particle
Greenberger-Horne-Zeilinger (GHZ) states and two-particle
Bell states. Apart from quantum sharing of classical
secrets, the idea has also been generalized to the sharing
of secret quantum information
\cite{Hillery-99,Karlsson-99,Cleve-99}, which is often
referred to as ``quantum state sharing" (also QSS). We
shall mainly be concerned with this notion of QSS in this
paper. Some recent theoretical works in this area can be
found in Refs. \cite{Gottesman00,Li-04,Deng-Zhou05,Hsu03,
Zhang-05,Schmid-05,Tyc-02}. On the experimental side, a
(2,3) threshold QSS protocol has been demonstrated in the
continuous variable regime \cite{Lance-04}. QSS using
pseudo-GHZ states has been reported earlier
\cite{Tittel-01}. Recently, a three-party QSS scheme has
also been demonstrated via four-photon entangled states
\cite{Chen-05}.

Controlled quantum teleportation (CQT) is an extension of
the original quantum teleportation protocol proposed by
Bennett \textit{et al.} \cite{Bennett93} in 1993. The idea
is to allow parties other than the receiver to have control
over the successful completion of a teleportation process.
In the first CQT protocol proposed by Karlsson \textit{et
al.} \cite{Karlsson-98}, an arbitrary single qubit state is
teleported to two receivers using a GHZ state, such that
only one of them can reconstruct the quantum state using
classical information provided by the other. Recently
quantum teleportation with multiparty control has also been
proposed \cite{Yang-05,Zhang-Man05,Deng-05}, in which the
receiver can fully recover the quantum state if all of the
controllers cooperate by communicating the outcomes of
their measurements to the receiver. Yang \textit{et al.}
\cite{Yang-05} and Zhang \textit{et al.} \cite{Zhang-Man05}
considered the controlled teleportation of a multiqubit
product state, and Deng \textit{et al.} \cite{Deng-05} a
two-qubit entangled state. Furthermore controlled
probabilistic teleportation of one- and two-qubit states
has also been studied lately \cite{Yan-03,Gao04}. Recent
experimental works on quantum teleportation can be found in
Refs. \cite{Houwelingen-06,Riedmatten-04,Pan-01,Kim-01}.

Most of the discrete variable QSS protocols proposed in the
literature are of the $(n,n)$-threshold type
\cite{Hillery-99,Karlsson-99,Li-04,Deng-Zhou05,Hsu03,
Zhang-05,Schmid-05}. A (2,3)-threshold scheme using qutrits
can be found in Ref. \cite{Cleve-99}. In the continuous
variable regime, general $(k,n)$-threshold schemes are
possible using optical interferometry \cite{Tyc-02,Tyc-03}.
Typically shares of quantum information are distributed by
teleportation or entanglement swapping via Bell or GHZ
states established between the sender and the players. In
ordinary QSS, after the completion of the distribution
process, the fate of the secret quantum information is
entirely left to the players. For example, in a
$(k,n)$-threshold scheme, any $k$ players may come together
anytime, extract the quantum secret and use it to execute a
certain task. This situation may not be desirable in many
real world situations, especially when it is crucial that
the extraction of the secret information (or the initiation
of the subsequent actions) requires authorization from
higher offices. For such cases, it is desirable to have a
secret sharing protocol where it is impossible for the
players to extract the secret information (even if all of
them agree to cooperate) before obtaining authorization
from superiors which we shall call ``controllers".

In this paper, we propose a new protocol which may be
viewed as a hybrid of QSS and CQT. We consider the
following situation. The dealer Alice has a $N$-qubit state
$\ket{\Psi_{1...N}}$ which encrypts a secret quantum
information $\ket{\xi}$, and it is to be shared by $n\le N$
players (\sd B1,\ldots,\sd Bn) and $m\le 2N$ controllers
(\sd C1,\ldots,\sd Cm). $\ket{\Psi_{1...N}}$ can be the
encoding state of any secret sharing scheme, and we shall
not specify it explicitly here. After the shares are
properly distributed, successful reconstruction of the
quantum secret $\ket{\xi}$ depends on two conditions: (1)
At least $m^*$ controllers must agree to release the
classical information they hold ($m^*$ depends on how the
classical shares are distributed; see below), and (2) a set
of at least $k$ players must collaborate to perform a joint
operation on the qubits they possess. Such a protocol may
be termed ``controlled quantum secret sharing" (CQSS).

It is easy to see that CQSS reduces to ordinary QSS if all
the controllers make public the classical information they
hold. Therefore the encoding state $\ket{\Psi_{1...N}}$,
together with its access structure, must satisfy the
theorems on QSS obtained in Refs.
\cite{Cleve-99,Gottesman00}. The existence of controllers
in CQSS adds another dimension to ordinary QSS. In order to
reconstruct the secret quantum information in CQSS, it is
not sufficient that a minimum number of players agree to
cooperate$-$they must first obtain authorization from the
controllers. Note that the players and the controllers play
asymmetric roles in a CQSS scheme: Namely the controllers
hold no quantum shares, therefore their role is not to
reconstruct the quantum secret themselves, but to control
when it should be done and which players are assigned to do
it. In QSS, there could also be asymmetry between the power
of different players \cite{Cleve-99,Gottesman00}, and this
feature can be retained in CQSS in the access structure of
the encoding state $\ket{\Psi_{1...N}}$. CQSS protocols are
useful in secure quantum communication networks. They are
also useful in the real world situation where the players
are field agents responsible for carrying out a certain
task, using the secret information encrypted in
$\ket{\Psi_{1...N}}$, and the controllers are superiors who
decide if and when the task should be carried out and who
to do it.

Most proposals for multiparty QSS
\cite{Hillery-99,Karlsson-99,Cleve-99,Li-04,Deng-Zhou05}
and teleportation
\cite{Karlsson-98,Yang-05,Deng-05,Yan-03,Gao04} with
multiparty control require ancillary entangled states
and/or collective measurements involving three or more
qubits. In some cases the number of involved qubits
increases with the $N$ or $m$, making them difficult to
implement by current technologies. In contrast, the CQSS
scheme to be presented below requires only Bell-basis
measurements and ancillary Bell states which are much
easier to produce and purify.

The first step of the protocol is to divide
$\ket{\Psi_{1...N}}$ into $n$ equal shares and distribute
them to the $n$ players. This can be achieved by
entanglement swapping (or teleportation) as shown in Ref.
\cite{Cheung05}. The procedure is conceptually quite simple
and we reproduce it below. To begin with, we define the
four Bell states:
 \bea
 &&\ket{\phi^{\pm}_{\mu\nu}}=\frac{1}{\sqrt{2}}
 \Big(\ket{0_\mu} \ket{1_\nu}\pm\ket{1_\mu}
 \ket{0_\nu}\Big),\\
 &&\ket{\varphi^{\pm}_{\mu\nu}}=\frac{1}{\sqrt{2}}
 \Big(\ket{0_\mu} \ket{0_\nu}\pm\ket{1_\mu}
 \ket{1_\nu}\Big),
 \eea
where the singlet state $\ket{\phi^-_{\mu\nu}}$ is also
known as the Einstein-Podolsky-Rosen (EPR) state.  We
assume that the Alice shares at least one EPR
$\ket{\phi^-_{\mu_i\nu_i}}$ with each player \sd Bi , so
that the total number of EPR states shared between them is
$N$. Similarly she shares at least one EPR state
$\ket{\phi^-_{\alpha_i\beta_i}}$ with each controller \sd
Ci , and the total number is $2N$. It is to be understood
that qubits $\{\mu_i\}$ and $\{\alpha_i\}$ belong to Alice,
qubits $\{\nu_i\}$ and $\{\beta_i\}$ belong respectively to
the players and controllers. We first show how to
distribute (or teleport) qubit-1 of $\ket{\Psi_{1...N}}$ to
\sd B1, as we shall see the procedure can be easily
generalized to include other qubits. The state
$\ket{\Psi_{1...N}}$ can always be cast in the form,
 \beq
 \ket{\Psi_{1...N}}
 =a\ket{0_1}\ket{\Phi_{2...N}}
 +b\ket{1_1}\ket{\Phi'_{2...N}},
 \label{Psi}
 \eeq
where $|a|^2+|b|^2=1$, and $\ket{\Phi_{2...N}}$ and
$\ket{\Phi'_{2...N}}$ are normalized states of $(N-1)$
qubits. Then we can write the product of
$\ket{\Psi_{1...N}}$ and $\ket{\phi^-_{\mu_1\nu_1}}$ as
 \bea
 \ket{\Psi_{1...N}}\ket{\phi^-_{\mu_1\nu_1}}
 &&\!\!\!\!\!=\frac{1}{2}\Big[
 \ket{\varphi^+_{1\mu_1}}
 \Big(a\ket{1_{\nu_1}}\ket{\Phi_{2...N}}
 -b\ket{0_{\nu_1}}\ket{\Phi'_{2...N}}\Big)\nonumber\\*
 &&\;+\;\,\ket{\varphi^-_{1\mu_1}}
 \Big(a\ket{1_{\nu_1}}\ket{\Phi_{2...N}}
 +b\ket{0_{\nu_1}}\ket{\Phi'_{2...N}}\Big)\nonumber\\*
 &&\;-\;\,\ket{\phi^+_{1\mu_1}}
 \Big(a\ket{0_{\nu_1}}\ket{\Phi_{2...N}}
 -b\ket{1_{\nu_1}}\ket{\Phi'_{2...N}}\Big)\nonumber\\*
 &&\;-\;\,\ket{\phi^-_{1\mu_1}}
 \Big(a\ket{0_{\nu_1}}\ket{\Phi_{2...N}}
 +b\ket{1_{\nu_1}}\ket{\Phi'_{2...N}}\Big)\!\Big].
 \label{product}
 \eea
Therefore a Bell-basis measurement by Alice on the pair of
qubits $(1,\mu_1)$ will entangle \sd B1's qubit-$\nu_1$ to
the inactive group $(2,\ldots,N)$, such that the resulting
$N$-qubit state depends on the outcome of Alice's
measurement. Comparing with Eqs. (\ref{Psi}), we see that
if Alice informs \sd B1 of the outcome, then by a local
unitary transformation on qubit-$\nu_1$, \sd B1 can rotate
the state of the $N$-qubit group $(\nu_1,2,\ldots,N)$ to
$\ket{\Psi_{\nu_1,2,\ldots,N}}$, which is exactly what we
started out with except that Alice's qubit-1 has been
replaced by \sd B1's qubit-$\nu_1$. The required unitary
operators for the four possible outcomes
($\ket{\varphi^+},\ket{\varphi^-},\ket{\phi^+}$,
$\ket{\phi^-})$ are respectively ($\sigma_z\sigma_x$,
$\sigma_x$, $\sigma_z$, $I$), where
 \bea
 \sigma_z\!&=&\!\ket{0}\bra{0}-\ket{1}\bra{1},\\
 \sigma_x\!&=&\!\ket{0}\bra{1}+\ket{1}\bra{0},\\
 I\,&=&\!\ket{0}\bra{0}+\ket{1}\bra{1}.
 \eea
Notice that the above procedure is entirely general, in the
sense that it is independent of the state of the inactive
qubits $(2,\ldots,N)$. Accordingly it can be repeated on
the other qubits until all of them are distributed to the
players (\sd B1,\ldots,\sd Bn).

Upon completing the distribution process, Alice would have
made $N$ Bell-basis measurements (one for each qubit in
$\ket{\Psi_{1...N}}$), and obtained $N$ Bell states,
$\{\ket{\psi^1},\ldots,\ket{\psi^N}\}$, where
$\ket{\psi^i}\in\{\ket{\phi^-},\ket{\phi^+},
\ket{\varphi^-},\ket{\varphi^+}\}$. Hence there is a
one-to-one correspondence between the qubits in
$\ket{\Psi_{1...N}}$ and the Bell states in the list
${\ket{\psi^1},\ldots,\ket{\psi^N}}$. Each Bell state
corresponds to two bits of classical information; for
instance we may assign
 \begin{subequations}
 \label{twobits}
 \bea
 &&\ket{\phi^-}\rightarrow 00,
 \quad\ket{\phi^+}\rightarrow 01,\\
 &&\ket{\varphi^-}\rightarrow 10,
 \quad\ket{\varphi^+}\rightarrow 11.
 \eea
 \end{subequations}
In an ordinary QSS scheme, the Bell-state information is
given to the players, so that they know what unitary
transformations to apply to their qubits in order to
recover the original state $\ket{\Psi}$. In contrast, for
the CQSS protocol being considered here, Alice distributes
the Bell-state information to the controllers (\sd
C1,\ldots,\sd Cm) instead. She could do it quantum
mechanically by teleporting the Bell states to the
controllers using the same method described above. However
it is simpler to just send the corresponding classical
information as follows. Suppose Alice wants to send two
bits $(x,y)$ to \sd Ci with whom she shares a pair of EPR
states $\ket{\phi^-_{\alpha_i\beta_i}}$ and
$\ket{\phi^-_{\alpha'_i\beta'_i}}$. From
 \bea
 \ket{\phi^-_{\alpha_i\beta_i}}\ket{\phi^-_{\alpha'_i\beta'_i}}
 =&&\!\!\!\!\!\!\frac{1}{2}\Big(
 \ket{\varphi^+_{\alpha_i\alpha'_i}}
 \ket{\varphi^+_{\beta_i\beta'_i}}
 -\ket{\varphi^-_{\alpha_i\alpha'_i}}
 \ket{\varphi^-_{\beta_i\beta'_i}}\nonumber\\*
 &&\!\!\!-
 \ket{\phi^+_{\alpha_i\alpha'_i}}\ket{\phi^+_{\beta_i\beta'_i}}
 +\ket{\phi^-_{\alpha_i\alpha'_i}}\ket{\phi^-_{\beta_i\beta'_i}}
 \,\Big),
 \eea
we see that a Bell-basis measurement by Alice on qubits
$\alpha_i$ and $\alpha'_i$ will leave the
$(\beta_i,\beta'_i)$ pair in one of the Bell states on \sd
Ci's side. Moreover the resulting Bell states on both sides
are random but identical. Hence by the convention given in
Eq. (\ref{twobits}), Alice and \sd Ci can obtain a pair of
random bits $(x',y')$ by independently performing a
Bell-basis measurement on qubits $(\alpha_i,\alpha'_i)$ and
$(\beta_i,\beta'_i)$ respectively. After that Alice can
announce the two-bit information $(x\oplus x',y\oplus y')$
(modulo 2) over a public channel, and \sd Ci will be able
to decode the secret bits $(x,y)$ since he knows $(x',y')$.

It is interesting to note that, instead of giving the
two-bit information of a Bell state to one controller as
described above, Alice could also choose to split the Bell
state and have it shared by two controllers. Again this can
be done by following the same procedure described earlier
for the distribution of $\ket{\Psi_{1...N}}$. In this case,
the two controllers involved must cooperate to make a join
Bell-basis measurement on their qubits in order to identify
the Bell state they share. Moreover if one of the
controllers does not cooperate, then the other one can get
absolutely no information about the Bell state
\footnote{Alice could also choose to do it classically by
splitting the two-bit information instead. However, in this
case, even if one of controllers does not cooperate, the
other still knows something about the Bell state.}.
Consequently, each individual controller has no complete
control over the release of the corresponding two-bit
information. This option may be useful in circumstances
where some controllers are of lower rank than others. This
concludes the specification of our CQSS protocol.

Consider the simplest case where each player receives one
qubit, and each controller one set of two-bit Bell-state
information ($i.e.,~ n=m=N$). If any $k$ controllers
release their two-bit information, then the $k$ players
holding the corresponding qubits can collaborate to extract
the secret information $\ket{\xi}$. Hence the minimum
number of consenting controllers is $m^*=k$ in this case.
Obviously if all the controllers agree to release the
information they hold, then any authorized set of $k$
players can extract the secret. If one of the controllers
withholds his two-bit information, then knowledge about the
corresponding qubit is completely hidden from the players.
This can be seen from Eq. (\ref{product}). Let the two-bit
information corresponding to the distribution of qubit-1 be
withheld, then the state of the $N$ qubits
$(\nu_1,2,\ldots,N)$ is an equal mixture of the four
possible outcomes as shown in Eq. (\ref{product}). The
corresponding density matrix is given by
 \bea
 \rho'_{{}_N}\!=&&\!\!\!\!\!\!
 \frac{1}{2}\,I_{\nu_1}\Big(
 |a|^2\ket{\Phi_{2...N}}\bra{\Phi_{2...N}}
 +|b|^2\ket{\Phi'_{2...N}}
 \bra{\Phi'_{2...N}}\Big),\nonumber\\
 =&&\!\!\!\!\!\!
 \frac{1}{2}\,I_{\nu_1}\text{Tr}_1
 \ket{\Psi_{1...N}}\bra{\Psi_{1...N}},
 \eea
where $I_{\nu_1}$ is the identity matrix for qubit-$\nu_1$.
Clearly $\rho'_{{}_N}$ contains absolutely no information
about qubit-1 in the original state $\ket{\Psi_{1...N}}$.
Hence if more than $(n-k)$ controllers withhold their
information, the quantum secret $\ket{\xi}$ is sealed, even
if all the players agree to cooperate.

In other situations where $n=N$ but $m<N$, some controllers
may receive more than one set of Bell-state information
(two bits). Then a controller holding more than $(n-k)$
sets would have veto power over the recovery of
$\ket{\xi}$. It follows that if everyone receives more than
$(n-k)$ sets, the recovery of $\ket{\xi}$ would require
unanimous consent from all the controllers. In the special
case where Alice keeps all the Bell-state information home,
then she becomes the sole controller who can decide not
only when to extract the secret information, but also which
players are assigned to do so. Finally, if Alice discloses
all the Bell state information to the players, then the
result is an ordinary QSS scheme.

If $n=1$, then CQSS reduces to the controlled teleportation
of $\ket{\Psi_{1...N}}$ with $m$ controllers. Recently two
CQT schemes with multiparty control have been proposed by
Yang \textit{et al.} \cite{Yang-05} and Zhang \textit{et
al.} \cite{Zhang-Man05}. Both schemes considered only the
controlled teleportation of a product state of $N$ qubits,
whereas our scheme can teleport an arbitrary $N$-qubit
entangled state. The scheme of Ref. \cite{Yang-05} requires
an ancillary multiqubit entangled state, which is difficult
if not impossible to implement when the number of qubits or
controllers becomes large. Ref. \cite{Zhang-Man05} also
employs only ancillary Bell states, and Alice transmits her
measurement results to the controllers via a public channel
using pre-established secret keys. In our case Alice could
split a Bell state and have it shared by two different
controllers; this option may be useful in certain
circumstances, but it is not available in Refs.
\cite{Yang-05,Zhang-Man05} or other CQT schemes.

As with all entanglement based quantum protocols, the
security of our scheme depends crucially on the quality of
the quantum entanglement connections between Alice and the
receivers (players and controllers). An important advantage
of this type of schemes is that the set-up, purification,
and checking of the shared EPR states can all be done prior
to and independent of the scheme itself
\cite{Yang-99,Bennett-96,BBPS96,Deutsch-96}. In the CQSS
protocol being considered here, Alice can conduct
additional security checking during the distribution
process by randomly inserting a number of decoy states
$\ket{\theta_j}$, with $\ket{\theta_j}\in
\{\ket{0},\ket{1},\ket{+x},\ket{-x}\}$ for example. With
$M$ decoy qubits, the state to be distributed becomes
 \beq
 \ket{\Psi'_{1...N+M}}=\ket{\Psi_{1...N}}
 \prod^M_{j=1}\ket{\theta_j},
 \eeq
At the conclusion of the distribution process, Alice
identifies the decoy qubits and asks the players to measure
them in appropriate bases and report the results. Checking
against her own record for discrepancies, Alice can detect
the existence of eavesdroppers. With appropriate number of
decoys, the chance of an eavesdropper escaping detection
can be made as small as desired. Similarly Alice can check
for eavesdropping activities between herself and the
controllers.

From the above discussions, it is clear that in order to
implement our CQSS protocol, the sender Alice must first
share $N$ EPR states with the $n\le N$ players, and $2N$
EPR states with the $m\le 2N$ controllers. To distribute
the the quantum shares to the players, Alice needs to make
$N$ Bell-basis measurements. A maximum of $2N$ Bell-basis
measurements are required to send the Bell state
information to the controllers ($N$ measurements if she
sends through classical channels only). Therefore, apart
from classical communications, the whole distribution
process requires $3N$ shared EPR states and $2N-3N$
Bell-basis measurements. Needless to say, extra resources
would be needed if decoy qubits are employed. Typically,
each decoy qubit would add one EPR state and one Bell-basis
measurement to the resources requirement given above.

In summary we have presented in this paper a new protocol
which we call ``controlled quantum secret sharing (CQSS)".
In this protocol, the encoding state $\ket{\Psi_{1...N}}$
is shared by $n$ players and $m$ controllers. After the
completion of the distribution process, further action is
to be initiated by the controllers by disclosing the
classical information they hold, then an authorized set of
players can proceed to extract the secret information as in
ordinary secret sharing schemes. We recap the procedure as
follows:

\begin{enumerate}
\item The dealer Alice possesses a $N$-qubit state
$\ket{\Psi_{1...N}}$ which encodes a secret quantum
information $\ket{\xi}$. In addition she shares a total of
$N$ EPR states with $n\le N$ players \{\sd B1,\ldots,\sd
Bn\}, and $2N$ EPR states with $m\le 2N$ controllers \{\sd
C1,\ldots,\sd Cm\}.
\item Alice divides $\ket{\Psi_{1...N}}$ into $n$ shares, and
distributes them to the players by entanglement swapping
(or teleportation) \cite{Cheung05}. In the process, she
obtains $N$ Bell states from the required Bell-basis
measurements. Each Bell state corresponds to two bits of
classical information. Normally Alice divides the Bell
state information into $m$ groups and transmits them to the
controllers through encrypted classical channels. However
if necessary Alice could also split any Bell state and have
it shared by two different controllers.
\end{enumerate}
In real world applications, the players may be considered
as field agents responsible for carrying out a task, using
the secret information encrypted in $\ket{\Psi_{1...N}}$,
and the controllers are superiors who decide if and when
the task should be carried out and who to do it. Our
protocol requires only ancillary EPR states and Bell-basis
measurements, so that it is relatively simple to implement.





\end{document}